%% file: main.tex
\documentclass[a4paper, 10pt, twocolumn]{article}

\input{gamma26-template}

\title{
    First Light with the Dark100 Array \\of PANOSETI Telescopes at \\Palomar Observatory 
}

\author{
	\authorstyle{
    N.~Korzoun\textsuperscript{\color{GammaDarkGreen}1},
    A.~Brown\textsuperscript{\color{GammaDarkGreen}2},
    S.~Fegan\textsuperscript{\color{GammaDarkGreen}3},
    L.~Fiedler\textsuperscript{\color{GammaDarkGreen}1},
    B.~Godfrey\textsuperscript{\color{GammaDarkGreen}4},
    J.~Holder\textsuperscript{\color{GammaDarkGreen}5},
    P.~Horowitz\textsuperscript{\color{GammaDarkGreen}6},
    A.W.~Howard\textsuperscript{\color{GammaDarkGreen}7},
    A.~Johnson\textsuperscript{\color{GammaDarkGreen}2},
    W.~Liu\textsuperscript{\color{GammaDarkGreen}4},
    J.~Maire\textsuperscript{\color{GammaDarkGreen}2},
    Y.~Popovych\textsuperscript{\color{GammaDarkGreen}1},
    E.~Pueschel\textsuperscript{\color{GammaDarkGreen}1},
    N.~Rault-Wang\textsuperscript{\color{GammaDarkGreen}4},
    S.~Ravikularaman\textsuperscript{\color{GammaDarkGreen}1},
    D.~Werthimer\textsuperscript{\color{GammaDarkGreen}4},
    S.A.~Wright\textsuperscript{\color{GammaDarkGreen}2}
    }
	\newline\newline
	{\footnotesize
        \textsuperscript{\color{GammaDarkGreen}1}\institution{Astronomisches Institut, Ruhr-Universität Bochum, Universitätsstr. 150, Bochum, 44801, Germany}\\
        \textsuperscript{\color{GammaDarkGreen}2}\institution{Department of Astronomy \& Astrophysics, University of California San Diego, 9500 Gilman Dr, La Jolla, 92093, USA}\\
        \textsuperscript{\color{GammaDarkGreen}3}\institution{Laboratoire Leprince-Ringuet, CNRS/IN2P3, École Polytechnique, Institut Polytechnique de Paris, Palaiseau, 91120, France}\\
        \textsuperscript{\color{GammaDarkGreen}4}\institution{Department of Astronomy, University of California Berkeley, 501 Campbell Hall, Berkeley, 94720, USA}\\
        \textsuperscript{\color{GammaDarkGreen}5}\institution{Department of Physics and Astronomy and the Bartol Research Institute, University of Delaware, 104 The Green, Newark, 19716, USA}\\
        \textsuperscript{\color{GammaDarkGreen}6}\institution{Department of Physics, Harvard University, 17 Oxford St, Cambridge, 02138, USA}\\
        \textsuperscript{\color{GammaDarkGreen}7}\institution{Department of Astronomy, California Institute of Technology, 1200 E. California Blvd., Pasadena, 91125, USA}
    }\\
}

\date{\today}
\newcommand{\website}{\href{https://plan.events.mpg.de/event/543/}{Proceedings of the 9th Heidelberg Gamma-Ray Symposium}} 

\begin{document}
\sansmath\maketitle
\thispagestyle{fancy}

\lettrineabstract{Dark100 is the first phase of an extensive array of Panoramic SETI (PANOSETI) telescopes at Palomar Observatory, California. PANOSETI telescopes are small, 0.5-m Fresnel lens telescopes instrumented with 1024-pixel silicon photomultiplier (SiPM) cameras. The telescopes were originally designed to search for ultra-fast optical transients, but they also function as cost-effective imaging atmospheric Cherenkov telescopes (IACTs). Dark100 will use seven of these widely separated IACTs to map the ultra-high-energy (UHE) gamma-ray sources on the Galactic Plane at high resolution. Dark100 will also leverage the IACT technique to search for evidence of ultra-heavy dark matter in a mass range that extends beyond 100 TeV. Deployment of the first four telescopes and enclosures was completed in December 2025, and commissioning of the next three is scheduled to finish by the end of 2026. The array has so far collected more than 100 hours of 3-telescope data on sources including the Crab Nebula and Markarian 421. Here we present the current status of Dark100.}

\begin{figure*}
    \centering
    \includegraphics[width=0.9\textwidth, keepaspectratio]{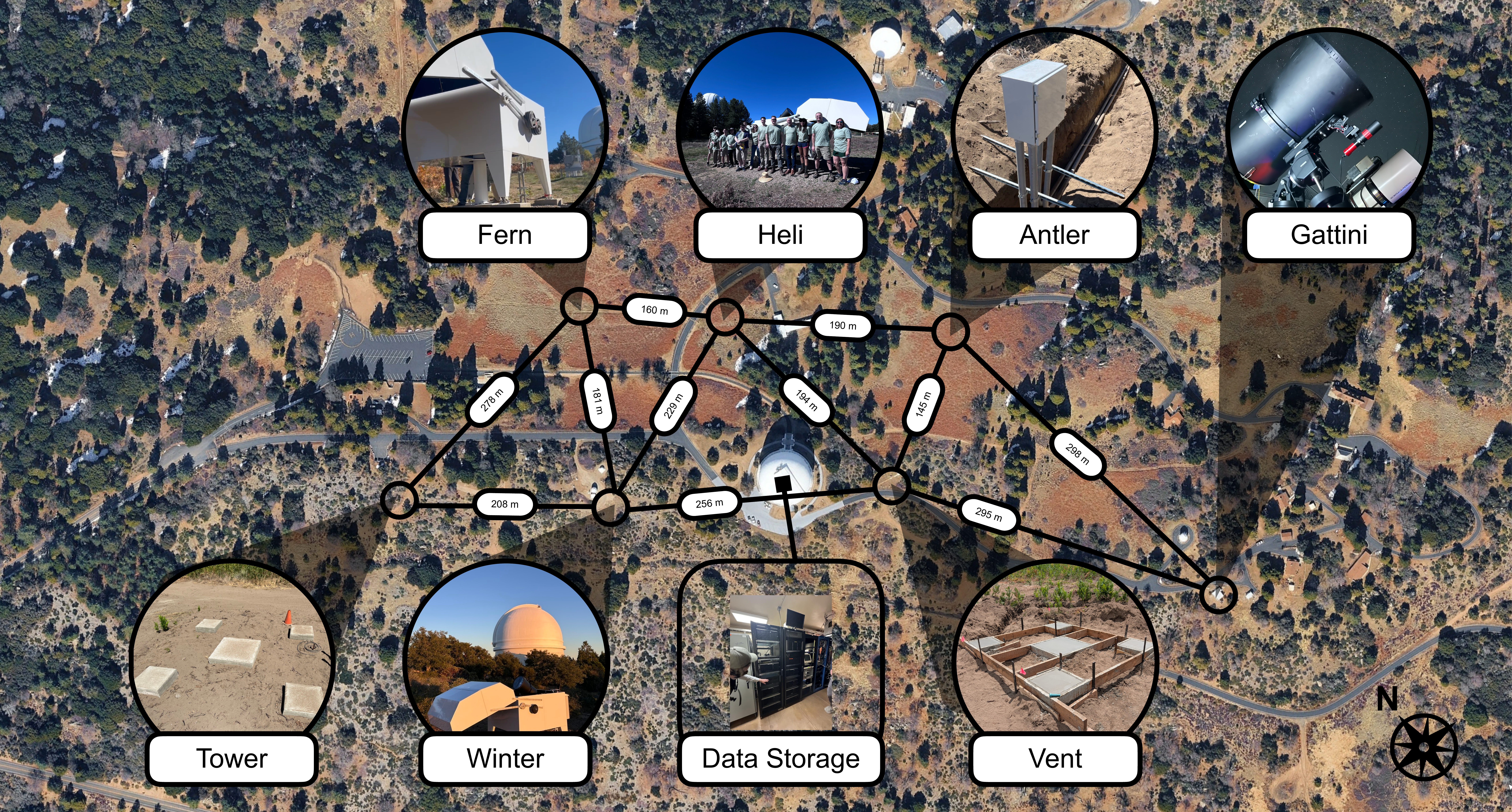}
    \caption{Schematic diagram of the Dark100 array of PANOSETI telescopes at Palomar Observatory. The Fern, Heli, Winter, and Gattini telescopes have completed commissioning. Construction is ongoing at the Antler, Tower, and Vent sites. Maps Data: Google, \copyright2025 Airbus.}
    \label{fig:array_layout}
\end{figure*}

\section{Introduction}
In recent years, the Large High Altitude Air Shower Observatory (LHAASO) has detected dozens of gamma-ray sources at ultra-high energies (UHE; $E>100$\,TeV), many with no sign of a spectral cutoff \cite{2024ApJS..271...25C}.
Gamma rays at these energies require parent particles with energies of order 1\,PeV, for which candidate sources have been dubbed "PeVatrons."
However, the angular resolution of particle detector arrays like LHAASO is limited, and many UHE sources are extended or have multiple plausible counterparts, so the origin of the emission often remains unclear.
Gamma-ray data are especially critical for the handful of objects that are not detected at any other frequencies.

Imaging atmospheric Cherenkov telescopes (IACTs) offer much finer angular resolution, but UHE observations pose two challenges for them.
First, the gamma-ray flux falls steeply with energy, so large effective areas and long exposures are required.
Second, UHE air showers are so extended that their images are often truncated by the edge of the camera's field of view (FoV), or they land far outside a compact array where reconstruction becomes difficult.
Dark100 is a new array of IACTs that tackles both of these challenges by using cost-effective telescopes from the Panoramic SETI (PANOSETI) experiment.

PANOSETI was originally designed to search for and study ultra-fast ($\sim$\,ns) optical transients \cite{10.1117/12.3103639}, and this same nanosecond-scale timing resolution is what makes these telescopes sensitive to Cherenkov emission from atmospheric air showers.
Early tests with VERITAS \cite{2022SPIE12184E..8BM} and standalone deployments at Lick Observatory \cite{2025icrc.confE.713K} demonstrated the potential of building a dedicated array of PANOSETI telescopes for gamma-ray astronomy.

The Dark100 array is located at Palomar Observatory in California.
In addition to studying ultra-fast optical transients, the array will survey the Galactic Plane for UHE gamma-ray emission at high resolution.
The name "Dark100" refers to another goal of the array: the search for ultra-heavy dark matter (particle masses above 100\,TeV).
For this purpose, Dark100 will use the same gamma-ray detection method, with photons as a tracer for potential dark matter interactions.

\section{Telescopes and Array Layout}
The optics, mechanics, and electronics of PANOSETI telescopes are described in \cite{2018SPIE10702E..5LM}, \cite{2021arXiv211112771B}, and \cite{2020SPIE11447E..7GL}.
In summary, these telescopes are much smaller than a typical IACT.
Instead of reflecting light with mirrors, PANOSETI telescopes focus light through a 0.46\,m diameter Fresnel lens.
The camera consists of 1024 silicon photomultiplier (SiPM) pixels, each covering 0.32\textdegree$\times$0.32\textdegree\ of sky, for a total FoV of 10\textdegree$\times$10\textdegree.
A guide camera is mounted on the side of each telescope to assist with tracking.
The telescopes are even small enough to be permanently contained within a weather-resistant enclosure.
The enclosures are compact aluminum boxes with a roof that lifts open on two counterweighted arms.
The four legs of the enclosure and the central pier that supports the telescope are all fixed to cement pads.
The opening mechanism is motor-driven and remotely operable.
Calibration of the telescopes is described in another contribution to these proceedings \cite{2026Gamma..P}, and in \cite{10.1117/12.3104216}.

A total of 26 potential sites were scouted at Palomar Observatory.
With the help of Monte Carlo simulations, we conducted a cost-benefit analysis to determine the ideal location for the first few telescopes of the array.
Telescope spacing is critical for maximizing photon yield within a given Cherenkov light pool, but we also had to consider the cost of trenching for routing data and power cables.
It was also important to plan ahead for future expansion, so that the first telescopes would not be left geographically isolated from future sites.

\begin{figure}
    \centering
    \includegraphics[width=0.95\columnwidth, keepaspectratio]{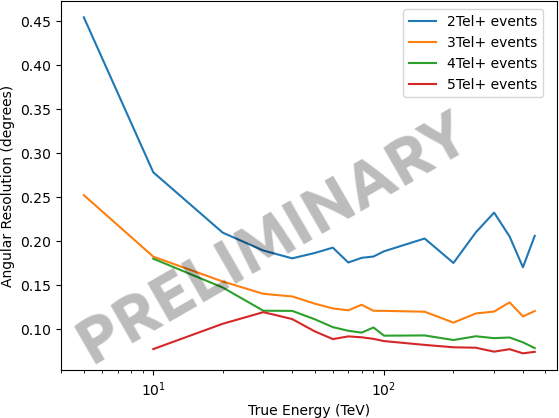}
    \caption{Simulated angular resolution for the full Dark100 array of seven PANOSETI telescopes. A range of telescope multiplicities are tested. Gamma/hadron selection cuts are not applied.}
    \label{fig:angular_res}
\end{figure}

\begin{figure}
    \centering
    \includegraphics[width=0.95\columnwidth, keepaspectratio]{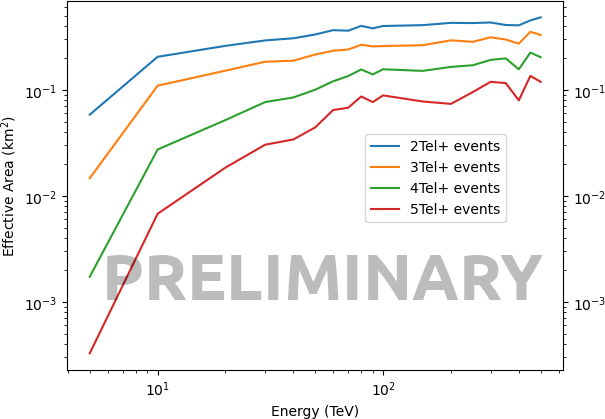}
    \caption{Simulated effective area for the full Dark100 array of seven PANOSETI telescopes. A range of telescope multiplicities are tested. Gamma/hadron selection cuts are not applied.}
    \label{fig:effA}
\end{figure}

A total of seven telescopes will be deployed by the end of 2026. 
The full array layout is shown in Figure \ref{fig:array_layout}.
The Gattini telescope was the first to be deployed, because a standalone enclosure was not necessary.
This telescope shares infrastructure with a preexisting telescope at the observatory.
The Fern, Heli, and Winter telescopes were commissioned next, with construction completing in December 2025.
The Gattini telescope is too far away from the other three to meaningfully contribute to coincident detection of air showers, but this will be remedied by future expansion of the array.
The Antler, Tower and Vent sites are currently being prepared.
Preliminary simulations of the angular resolution and effective area of the full Dark100 array are shown in Figures \ref{fig:angular_res} and \ref{fig:effA}.
For events triggering at least three telescopes, the simulated array reaches an angular resolution of $\sim$0.12\textdegree\ and an effective area of $\sim$0.3\,km$^2$ above 100\,TeV.

\section{Observing Strategy}
At this early stage of the project, Dark100's primary objective is to achieve a significant detection of the Crab Nebula.
Observations of the Crab Nebula, therefore, take priority whenever it is visible.
A 50-hour, three telescope dataset was collected at the earliest feasible opportunity, in January 2026.
The analysis of this dataset is the topic of the next section.

When the Crab Nebula is not visible, Dark100 is strategically targeting bright point-sources (e.g. flaring active galactic nuclei) and hard-spectrum UHE objects identified by LHAASO \cite{2024ApJS..271...25C}.
The current status of the Dark100 observing program is described in Table \ref{tab:sources}.
Not only are these sources among the brightest TeV objects in the sky, but in many cases, their spectra extend well into the UHE regime.
We therefore expect that these sources provide Dark100 with the simplest path toward establishing a catalog of significantly detected gamma-ray emitters. 

\begin{table}
    \centering
    \begin{tabular}{|p{3.4cm}|p{3.4cm}|}
        \hline
        \rowcolor{lightgray}
        \large{\textbf{Target}} & \large{\textbf{Exposure (hours)}} \\
        \hline \hline
        Crab Nebula & 134 \\
        \hline
        Mrk 421 & 71 \\
        \hline
        1ES 1959+650 & 30 \\
        \hline
        MGRO J2019+37 & 20 \\
        \hline
        Mrk 501 & 8 \\
        \hline
        Sgr A* & 5 \\
        \hline
    \end{tabular}
    \caption{List of gamma-ray targets observed to date with Dark100. Note that this list excludes serendipitous observing of other sources contained within the 10 degree FoV of a PANOSETI telescope. Exposures are summed over the full observing period and were not necessarily taken with all telescopes simultaneously.}
    \label{tab:sources}
\end{table}

Looking ahead, Dark100 plans to map UHE sources on the Galactic Plane at high resolution.
PANOSETI telescopes are particularly well-suited for this goal because of the 100 square degree FoV of the camera, which allows multiple sources to be observed simultaneously.
Consider, for example, the Cygnus region.
When PANOSETI telescopes target MGRO J2019+37, there are actually three additional bright TeV objects contained in the same field (the supernova remnant Gamma Cygni, the pulsar wind nebula CTB 87, and the first unidentified TeV source TeV J2032+4130) \cite{2018ApJ...861..134A}.

Dark100 will also observe dark matter rich targets like dwarf spheroidal galaxies and the Galactic Center, which doubles as a promising region for PeVatron studies.
As an early test for the observing conditions at low elevation, we collected 5 hours of data on Sgr A*.
Prospects for dark matter searches with Dark100 are reported on separately in these proceedings \cite{2026Gamma..R}.

\begin{figure*}
    \centering
    \includegraphics[width=0.9\textwidth, keepaspectratio]{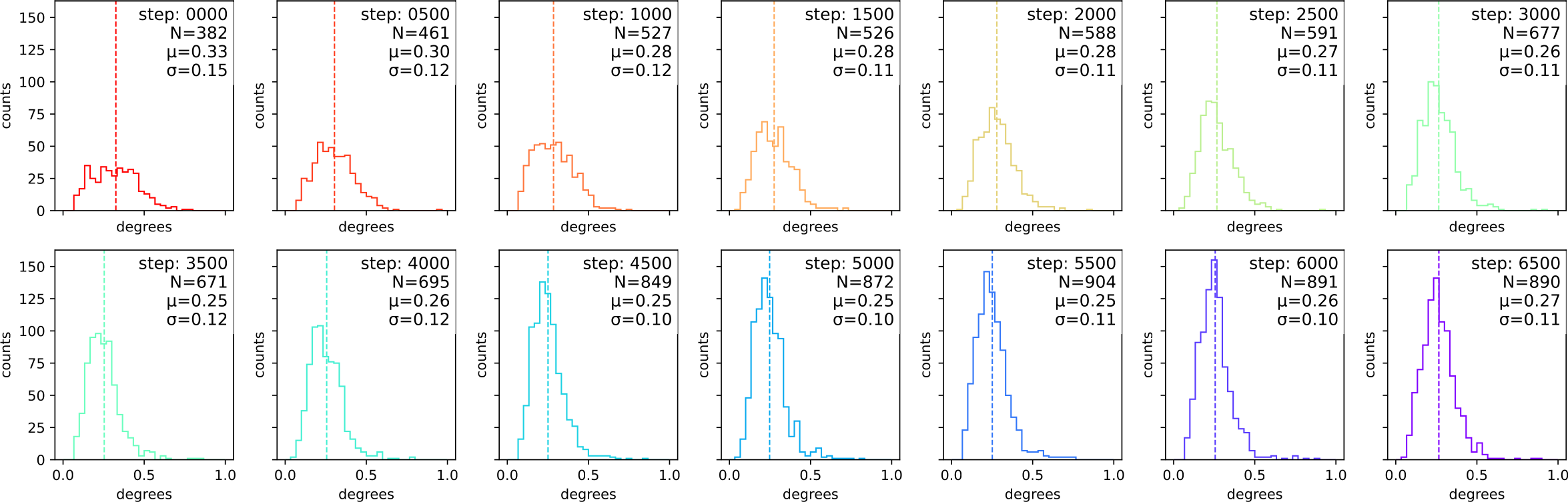}
    \caption{Distribution of image widths at different camera focal positions. The camera focuser was adjusted from "red-focus" to "blue-focus" in increments of 500 steps (2\,mm), and was exposed for 10 minutes at each increment. The total number of recorded events ($N$), the distribution mean ($\mu$), and standard deviation ($\sigma$) are listed for each tested focal position.}
    \label{fig:widths}
\end{figure*}

\section{Early Data}
The Fern, Heli, and Winter telescopes completed installation in late December 2025.
These telescopes recorded their first data on the Crab Nebula, taken in January 2026 and amounting to roughly 50 hours.
Using this dataset, we identified several aspects of the telescope configuration that needed improvement and made adjustments to optimize performance for future observations.
Since this dataset predates those adjustments, calibrating it was a nuanced task.

Firstly, the absolute pointing of each telescope needed correcting.
The positions of stars were identified both in guide camera images, and in the pedestal variances of the recorded data.
In most cases, the corrections were within the size of a pixel.

We also noticed that the median camera trigger rate was about half of what was measured with the pathfinder array at Lick Observatory \cite{2025icrc.confE.713K} (0.5\,Hz vs. 1\,Hz), despite using an identical 6.5 photoelectron trigger threshold.
Related to this, the distributions of image parameters varied for each telescope.
Suspecting the telescopes were not all perfectly focused, we performed a small experiment with the Fern telescope.
We pointed the telescope to 80\textdegree\ elevation and deliberately moved the camera out of focus.
The camera focuser was adjusted in increments of 500 steps (2\,mm), and was exposed for 10 minutes at each increment.
The resulting distributions of the image widths are shown in Figure \ref{fig:widths}.
The location of the focal plane has a significant impact on the width distributions due to chromatic aberration of the Fresnel lens.
Shifting the focal plane by just 20\,mm more than doubled the trigger rate, lowered the mean width by 0.08 degrees, and lowered the standard deviation by 0.05 degrees.
By examining their width distributions and checking the focuser positions, we determined that the Fern and Winter telescopes were not optimally focused while collecting data on the Crab Nebula.
All telescopes have since been refocused using this width-scanning method, which optimizes the focus for light with a Cherenkov spectrum.

Lastly, the pedestal variances showed significant structure across each camera, rather than our expectation of a flat measurement of a roughly uniform and steady light source (night sky background).
This resulted from a hardware issue that caused pixel gains to not be correctly set.
There was a break in the logic of the high voltage loop, which normally sends a command to update pixel voltages after measuring them.
The voltage measurements were correct, but the command to update the voltages did not execute.
This bug has since been fixed, and pedestal variances are now much flatter across each camera.

The Gamma/hadron cuts used in this analysis were derived from Monte Carlo simulations.
To ensure there were enough pixels to accurately parameterize each event, we required each image to retain at least three pixels after cleaning.
Then we apply a cut at $\theta^2 < 0.04$ deg$^2$ and a cut on image centroid distance $< 2$\textdegree.
To account for the differences in focus, we also devised a width cut that was unique for each telescope, requiring (\textit{width} $ < \mu + 0.5\,\sigma$), where $\mu$ is the mean of a telescope's width distribution across the full dataset, and $\sigma$ its standard deviation.

\begin{figure}
    \centering
    \includegraphics[width=0.95\columnwidth, keepaspectratio]{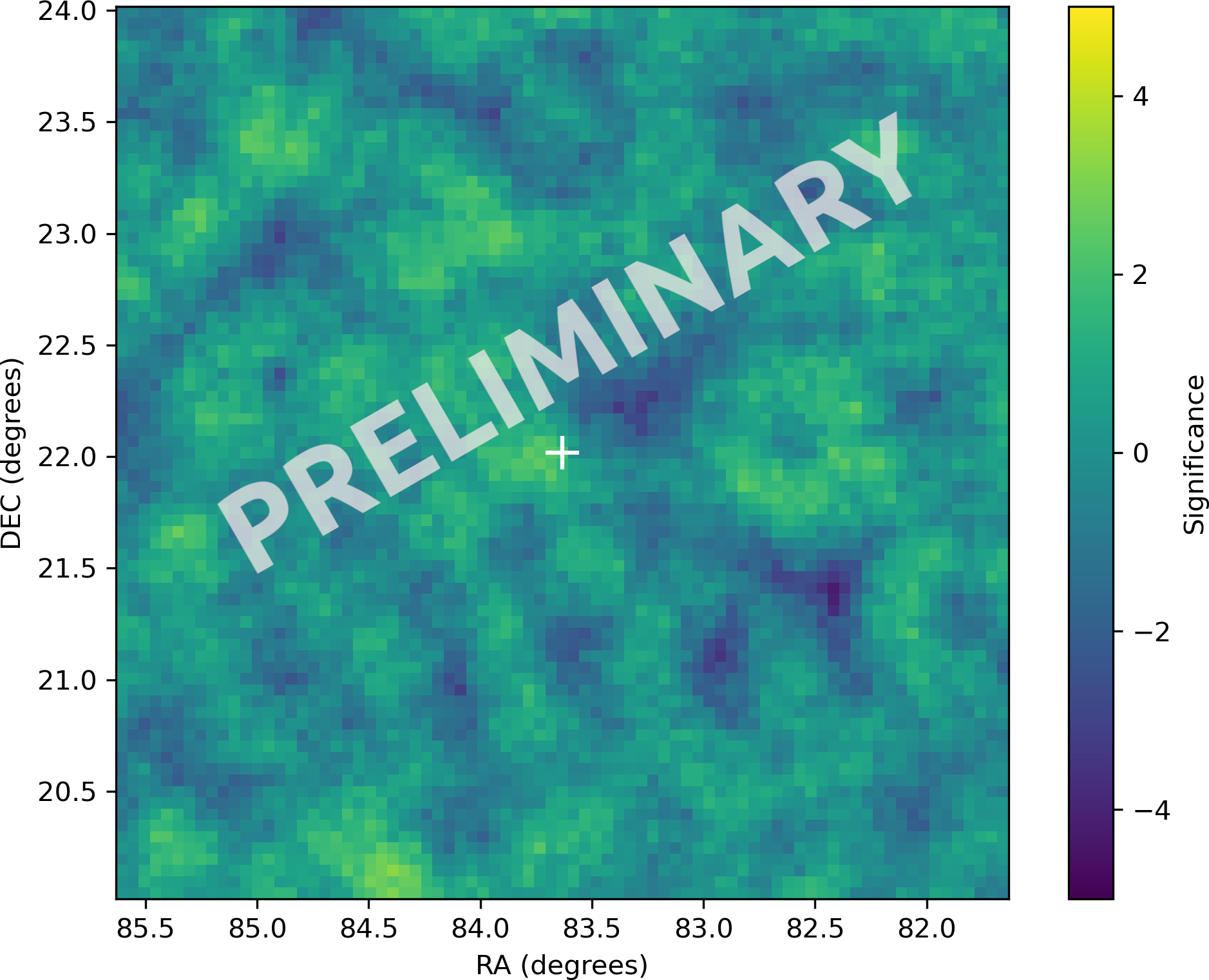}
    \caption{Significance map of the 50 hours, 3-telescope Crab Nebula dataset. The position of the Crab Nebula is marked with a white '$+$'. At the position of the Crab Nebula, the measured significance is 2.03 $\sigma$.}
    \label{fig:sigmap}
\end{figure}

\begin{figure}
    \centering
    \includegraphics[width=0.95\columnwidth, keepaspectratio]{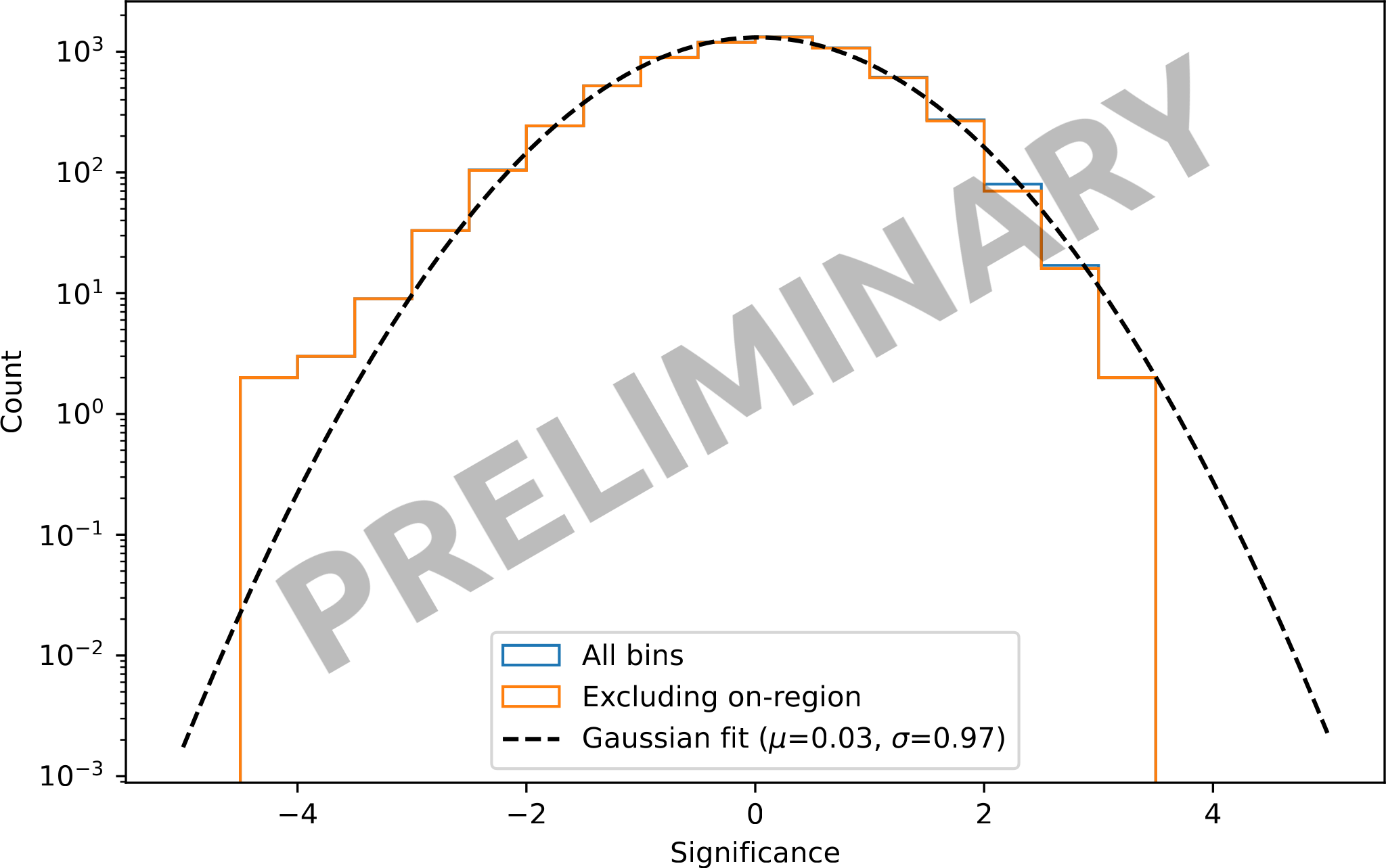}
    \caption{Distribution of significances shown in the map in Figure \ref{fig:sigmap}.}
    \label{fig:sigdist}
\end{figure}

Likely as a consequence of the defocusing and incorrect pixel gains, analysis of the full 50-hour Crab Nebula dataset did not yield a detection.
A significance map of the Crab Nebula region is shown in Figure \ref{fig:sigmap}.
At the position of the Crab Nebula, the measured significance is 2.03 $\sigma$.
There is no evidence for significant emission in any region of Figure \ref{fig:sigmap}, also seen in the distribution of significances in Figure \ref{fig:sigdist}. 

\begin{figure}
    \centering
    \includegraphics[width=0.9\columnwidth]{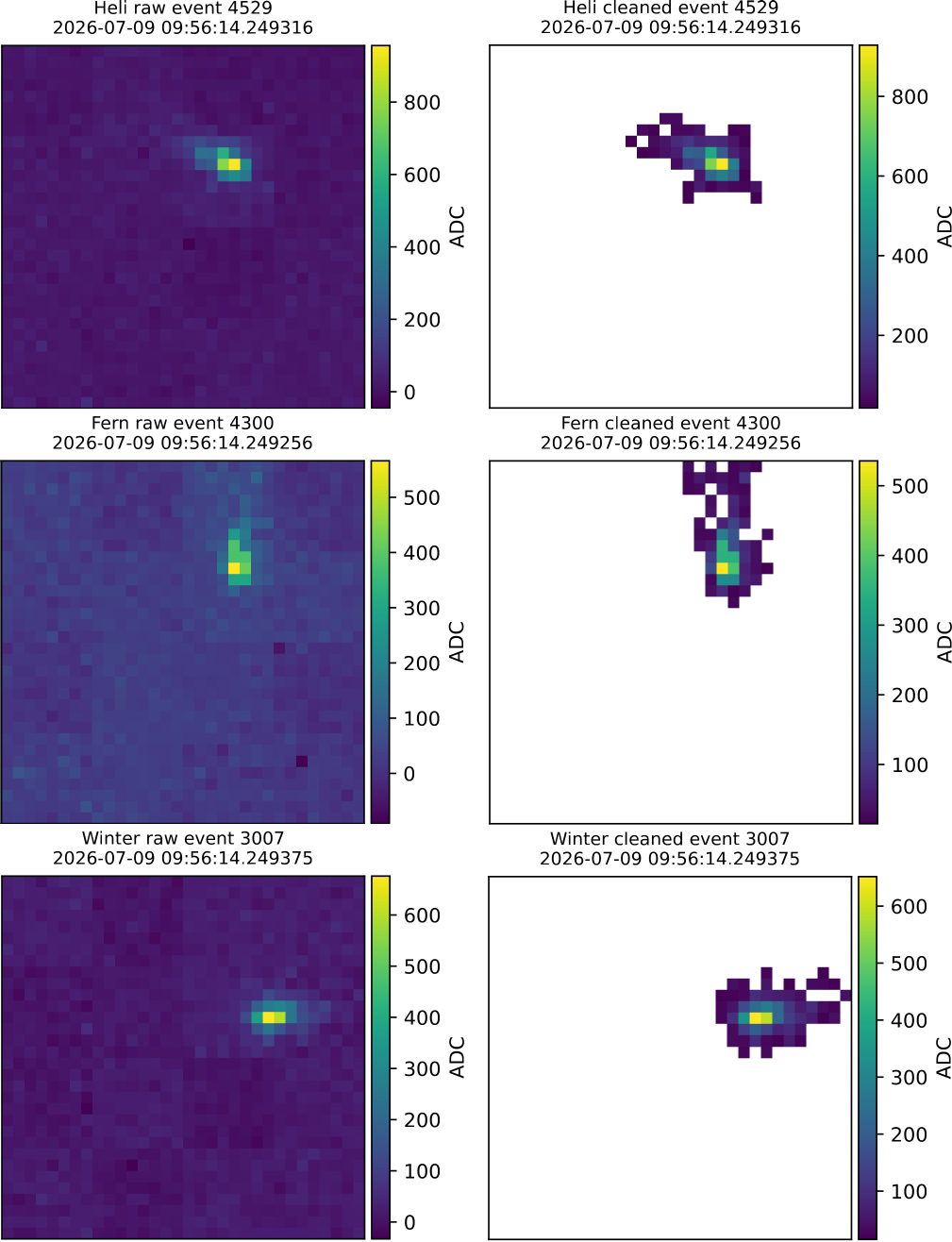}
    \caption{An example of an event recorded in the latest July 2026 dataset which triggered 3 PANOSETI telescopes in coincidence. The left column shows the raw images recorded by each telescope. The right column shows the same event after image cleaning.}
    \label{fig:event}
\end{figure}

The January 2026 observations therefore served primarily as commissioning data, and all of the issues identified in them have since been corrected.
Data collected since June 2026 show major improvements in quality.
An example event, which was seen in coincidence between three telescopes, is shown in Figure \ref{fig:event}.

\section{Discussion and Future Work}
Data collected on the Crab Nebula were imperfect, which is to be expected for observations taken so soon after the telescopes had been commissioned.
These challenges, coupled with PANOSETI's estimated energy threshold of a few TeV, make a nondetection of the Crab Nebula in 50 hours an unsurprising result.
Dark100 is constructing three additional PANOSETI telescopes, which will reduce the exposure time required.
Several technical issues in data taking have been addressed, improving the prospects for a Crab Nebula detection in future data.

The PANOSETI gamma-ray analysis pipeline is in the process of being upgraded to use the community standard tools sim\_telarray \cite{2008APh....30..149B}, ctapipe \cite{2024icrc.confE.703C}, and Gammapy \cite{2023A&A...678A.157D}.
Compatibility with these tools will provide better accessibility to Dark100 data, which we plan to make available in a public archive in the future.
Integrating these tools with Dark100 will also enable joint simulation and data analysis with other observatories.

With all seven telescopes commissioned and the pipeline upgrades in place, Dark100 will be well-positioned to map UHE gamma-ray emission across the Galactic Plane at high resolution and to search for signs of ultra-heavy dark matter, extending the scientific reach of PANOSETI well beyond its original design as a transient-search instrument.

\section*{Acknowledgments}
The PANOSETI program is a collaborative initiative involving researchers and instrumentation teams from institutions including UC San Diego, UC Berkeley, Harvard University, Caltech, Ruhr-Universität Bochum, and the University of Delaware. It is made possible through the generous support of Franklin Antonio and the Bloomfield Family Foundation. Additional support is provided by The Planetary Society and the Bosack/Kruger Foundation (Harvard SETI), NSF grant 1407804 and the Marilyn \& Watson Alberts SETI Chair fund (UC Berkeley), the ERC Consolidator Grant No. 101124914 (Ruhr-Universität Bochum), and the General University Research Program (University of Delaware). We gratefully acknowledge the staff and programmatic support from the California Institute of Technology (Palomar Observatory).

\printbibliography[title={References}]

%% Full authors list for collaborations (moved into \author block above)
% \clearpage
% \onecolumn
% \section*{Full Author List: PANOSETI Collaboration}
% \scriptsize
% \noindent
% \vskip1cm
%
% % Author list here
% A. Brown$^{1}$,
% S. Fegan$^{2}$,
% L. Fiedler$^{3}$,
% B. Godfrey$^{4}$,
% J. Holder$^{5}$,
% P. Horowitz$^{6}$,
% A.W. Howard$^{7}$,
% A. Johnson$^{1}$,
% N. Korzoun$^{3}$,
% W. Liu$^{4}$,
% J. Maire$^{1}$,
% Y. Popovych$^{3}$,
% E. Pueschel$^{3}$,
% N. Rault-Wang$^{4}$,
% S. Ravikularaman$^{3}$,
% D. Werthimer$^{4}$,
% S.A. Wright$^{1}$
%
% \vskip1cm
%
%
% $^{1}$Department of Astronomy \& Astrophysics, University of California San Diego, 9500 Gilman Dr, La Jolla, 92093, USA
% $^{2}$Laboratoire Leprince-Ringuet, CNRS/IN2P3, École Polytechnique, Institut Polytechnique de Paris, Palaiseau, 91120, France
% $^{3}$Astronomisches Institut, Ruhr-Universität Bochum, Universitätsstr. 150, Bochum, 44801, Germany
% $^{4}$Department of Astronomy, University of California Berkeley, 501 Campbell Hall, Berkeley, 94720, USA
% $^{5}$Department of Physics and Astronomy and the Bartol Research Institute, University of Delaware, 104 The Green, Newark, 19716, USA
% $^{6}$Department of Physics, Harvard University, 17 Oxford St, Cambridge, 02138, USA
% $^{7}$Department of Astronomy, California Institute of Technology, 1200 E. California Blvd., Pasadena, 91125, USA

\end{document}

%% file: gamma26-template.tex
\usepackage[british]{babel}
\usepackage{microtype}
\usepackage{amsmath,amsfonts,amsthm}
\usepackage[svgnames,table]{xcolor}
\definecolor{GammaDarkGreen}{rgb}{0.113,0.278,0.2745}
\definecolor{GammaCharcoal}{rgb}{0.0706, 0.0706, 0.0706}
\definecolor{GammaRed}{HTML}{B71325}
\definecolor{GammaGreen}{HTML}{004846}
\usepackage{booktabs}
\usepackage{lastpage}
\usepackage{graphicx}
\usepackage{enumitem}
\setlist{leftmargin=*,itemsep=0.4ex,parsep=0ex,topsep=0.8ex,partopsep=0ex}
\setlist[itemize]{label=\color{GammaRed}$\triangleright$}
\usepackage{sectsty}
\chapterfont{\normalfont\Large\color{GammaRed}}
\sectionfont{\normalfont\Large\color{GammaRed}}
\subsectionfont{\normalfont\large\color{GammaCharcoal}}
\subsubsectionfont{\normalfont\normalsize\color{GammaGreen}}

\usepackage{subfig}

\usepackage{caption}
\DeclareCaptionLabelFormat{abbrevfig}{\color{GammaRed}Fig.\ #2}
\DeclareCaptionLabelFormat{abbrevtab}{\color{GammaRed}Tab.\ #2}
\usepackage[pdftex]{hyperref}
\pdftrailerid{}
\hypersetup{colorlinks=true,linkcolor=GammaDarkGreen,citecolor=GammaDarkGreen,filecolor=GammaDarkGreen,urlcolor=GammaDarkGreen}

\addto\extrasenglish{%
}

\usepackage{geometry}
\usepackage[utf8]{inputenc}
\usepackage[T1]{fontenc}
\IfFileExists{nimbusmononarrow.sty}{\usepackage{nimbusmononarrow}}{}
\usepackage{tgheros}
\usepackage{sansmath}
\usepackage{fancyhdr}
\usepackage{titling}

\newcommand{\authorstyle}[1]{{\large\color{GammaCharcoal}#1}}
\newcommand{\institution}[1]{{\footnotesize\itshape\color{GammaDarkGreen}#1\/}}

\pretitle{%
    \noindent
    \begin{minipage}[t]{0.63\textwidth}\vspace{20pt}
        \LARGE\color{GammaRed}
}

\posttitle{%
    \end{minipage}%
    \hfill
    \begin{minipage}[t]{0.33\textwidth}\vspace{-50pt}
        \raggedleft
        \includegraphics[width=\linewidth]{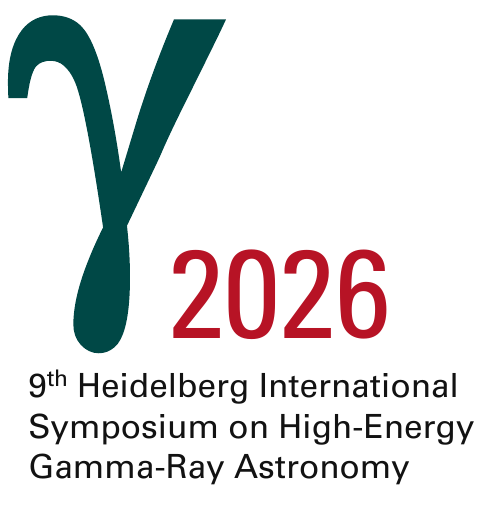}
    \end{minipage}%
}

\preauthor{\color{GammaDarkGreen}}
\postauthor{}
\predate{\begin{flushleft}\large\color{GammaRed}}
\postdate{\par\end{flushleft}}

\usepackage{lettrine}
\usepackage{xstring}
\newcommand{\initial}[1]{\lettrine[lines=2,findent=4pt,nindent=0pt]{\color{GammaRed}\emph{#1}}{}}
\newcommand{\lettrineabstract}[1]{\StrLeft{#1}{1}[\firstletter]\initial{\firstletter}\emph{{\StrGobbleLeft{#1}{1}}}}

\usepackage[backend=bibtex,style=numeric-comp,sorting=none,giveninits=true,citestyle=numeric]{biblatex}
\usepackage[autostyle=true]{csquotes}